\documentclass[superscriptaddress,twocolumn]{revtex4}
\usepackage{graphicx}
\usepackage{color}

\begin{document}

\title{Potential model for 
nuclear astrophysical fusion reactions with a square-well potential} 

\author{R. Ogura}
\affiliation{ 
Department of Physics, Tohoku University, Sendai 980-8578,  Japan} 

\author{K. Hagino}
\affiliation{ 
Department of Physics, Tohoku University, Sendai 980-8578,  Japan} 
\affiliation{Research Center for Electron Photon Science, Tohoku University, 1-2-1 Mikamine, Sendai 982-0826, Japan}

\author{C.A. Bertulani}
\affiliation{
Department of Physics and Astronomy, Texas A\&M University-Commerce, 
Commerce, Texas 75429-3011, USA}


\begin{abstract}
The potential model for 
nuclear astrophysical reactions requires a considerably shallow  nuclear potential 
when a square-well potential is employed to fit experimental data. 
We discuss the origin of this apparently different behavior from that 
obtained with a smooth Woods-Saxon potential, for which a deep potential is often employed. 
We argue that due to the sharp change of the potential at the boundary 
the radius parameter tends to be large in the square-well 
model, which results in a large absorption radius. 
The wave function then needs to be suppressed in the 
absorption region, which can eventually 
be achieved by using a shallow potential. We thus clarify the reason why the 
square-well potential has been able to reproduce a large amount of fusion data.
\end{abstract}


\maketitle

\section{Introduction}

Heavy-ion fusion reactions, such as $^{12}$C+$^{12}$C and $^{16}$O+$^{16}$O 
reactions, play an important role in nuclear astrophysics 
\cite{BTW85,BD04,TN09}. 
These reactions take place at extremely low energies, and a direct 
measurement of the reaction cross sections to obtain the astrophysical
fusion rates is almost impossible. 
It is therefore indispensable to extrapolate experimental data at 
higher energies down to the region which is relevant to 
nuclear astrophysics. For this purpose, the potential model with 
a Woods-Saxon potential has often been used. Alternatively, one can 
also use a square-well potential, as has been advocated very successfully by Michaud and 
Fowler \cite{MF70}. The fusion probability can be evaluated analytically 
with such a square-well potential, and 
the calculation becomes considerably simplified. 
See e.g., Refs. \cite{Tang12,Zhang16} for recent applications of the 
square-well model to the $^{12}$C+$^{12}$C and 
$^{12}$C+$^{13}$C reactions. 

Despite its simple nature, a square-well potential accounts for a large amount of
experimental data, sometimes even better than a fit with a Woods-Saxon 
potential \cite{BTW85}. However, it has been recognized that the resultant 
square-well potential, that is used for a total (the nuclear + the 
Coulomb) potential, has to be repulsive \cite{Fowler75,Dayras76}. 
For instance, for the $^{12}$C+$^{12}$C reaction, the best fit was 
obtained with the square-well potential, $V(r)=V_0\,\theta(R-r)~~(r\leq R)$, 
with $V_0=+5.8$ MeV and $R=7.50$ fm \cite{Fowler75}. Even though 
the value of $V_0$ is somewhat smaller than the Coulomb energy at $r=R$, 
that is, $V_c=6.9$ MeV, and thus the nuclear interaction is still 
attractive, the potential depth for the nuclear potential, $V_0-V_c$, 
is only 1.1 MeV, which is unusually small. The same tendency has been 
also found for the 
$^{12}$C+$^{16}$O and 
the $^{16}$O+$^{16}$O reactions \cite{Fowler75}. 

The purpose of this paper is to clarify the origin of a shallow 
depth of a square-well potential for nuclear astrophysical reactions. 
To this end, we shall study the sensitivity of fusion cross sections to the 
parameters of the square-well potential, such as the range of the imaginary 
part and the depth of the real part. 

\section{Square-well potential model} 

In the square-well potential model, one considers the 
following radial wave function for the relative motion 
between two nuclei \cite{BTW85}: 
\begin{eqnarray}
u(r)&=&Te^{-iKr} ~~~~~~~~~~~~~~~~~~~~~~~~(r < R), \label{bc1} \\
&=&H_l^{(-)}(kr)-S_lH_l^{(+)}(kr)~~~~(r\geq R), \label{bc2}
\end{eqnarray}
with $K=\sqrt{2\mu[E-(V_0-iW_0)]/\hbar^2}$ and 
$k=\sqrt{2\mu E/\hbar^2}$, $\mu$ and $E$ being the reduced mass 
and the incident energy in the center of mass frame, respectively. 
Here, $T$ and $S_l$ are the transmission and reflection coefficients, 
respectively, and $l$ is the partial wave. 
$H_l^{(+)}$ and $H_l^{(-)}$ are the outgoing and the incoming Coulomb 
wave functions, respectively, which are given in terms of the regular 
and the irregular Coulomb wave functions as 
$H_l^{(\pm)}=G_l\pm F_l$. 
The form of the wave function for $r<R$ is nothing but the incoming 
wave boundary condition \cite{HT12,ccfull}, which assumes a strong 
absorption in the region $r<R$. The imaginary part of the square well 
potential, $-iW_0$, allows an absorption even for $E<V_0$. 
From the matching condition of the wave function at $r=R$, 
one obtains \cite{MSV70}
\begin{equation}
1-|S_l|^2=\frac{4\frac{P_l}{KR}}
{\left(1+\frac{P_l}{KR}\right)^2+\left(\frac{s_l}{KR}\right)^2},
\end{equation}
with 
\begin{eqnarray}
P_l&=&\frac{kR}{F_l^2+G_l^2}, \label{P_l} \\
s_l&=&kR\,\frac{F_lF_l'+G_lG_l'}{F_l^2+G_l^2}, \label{s_l}
\end{eqnarray}
where the right hand side of Eqs. (\ref{P_l}) and (\ref{s_l}) are 
evaluated at $kR$. 
Fusion cross sections are then computed as \cite{HT12}, 
\begin{equation}
\sigma_{\rm fus}(E)=\frac{\pi}{k^2}\,\sum_l(2l+1)(1-|S_l|^2). 
\end{equation}
With those fusion cross sections, the astrophysical $S$-factor is defined as, 
\begin{equation}
S(E)=E\sigma_{\rm fus}(E)\,e^{2\pi\eta(E)},
\end{equation}
where $\eta(E)=Z_PZ_Te^2/(\hbar v)$ is the Sommerferd parameter. Here, 
$Z_P$ and 
$Z_T$ are the charge number of the projectile and the target nuclei, 
respectively, and $v=\sqrt{2E/\mu}$ is the velocity for the relative motion. 

For fusion of two identical bosons, such as $^{12}$C+$^{12}$C and 
$^{16}$O+$^{16}$O, one has to symmetrize the wave function with respect 
to the interchange of the two nuclei. The fusion cross sections are 
then evaluated as \cite{RH15}, 
\begin{equation}
\sigma_{\rm fus}(E)=\frac{\pi}{k^2}\,\sum_l(1+(-1)^l) (2l+1)(1-|S_l|^2). 
\end{equation}
In this case, only even partial waves contribute to fusion cross 
sections. 

\begin{figure} [tb]
\includegraphics[scale=0.6,clip]{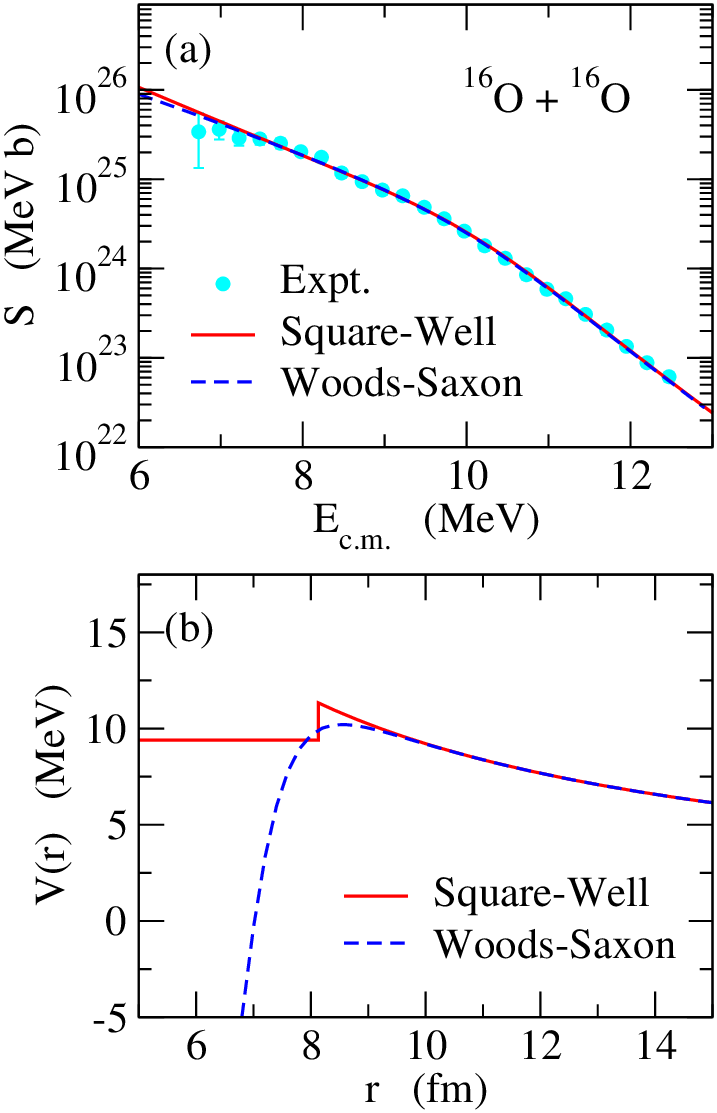}
\caption{
(The upper panel) The astrophysical $S$-factor for the 
$^{16}$O+$^{16}$O reaction. The solid and the dashed lines are 
obtained with a square-well and a Woods-Saxon potentials, respectively. 
The experimental data are taken from Ref. \cite{Thomas86}. 
(The lower panel) The radial dependence of the total 
potentials used in the calculations shown in the 
upper panel. 
}
\end{figure}

The upper panel of Fig. 1 shows the astrophysical $S$-factor for the 
$^{16}$O+$^{16}$O reaction. The solid line is obtained with a square-well 
potential with $V_0=9.4$ MeV, $R=8.13$ fm, and $W_0=2.1$ MeV. 
The reduced mass is taken to be $\mu=m(^{16}{\rm O})/2$, where 
$m(^{16}{\rm O})$ 
is the experimental mass for the $^{16}$O nucleus. 
For comparison, the figure also shows the result of the Woods-Saxon 
potential with the depth, the range, and the diffuseness parameters 
for the real part of 
$V_0=-54.5$ MeV, $R=6.5$ fm, and $a=0.45$ fm, 
respectively (the dashed line). The parameters for the imaginary 
part are taken to be $W_0=10.0$ MeV, $R_w=5$ fm, and $a_w=0.1$ fm. 
Those calculations are compared with the experimental data \cite{Thomas86}. 
One can see that both calculations reproduce the data equally well. 

The lower panel of the figure shows the radial dependence of the two 
potentials employed. Evidently, the square-well potential is much shallower than 
the Woods-Saxon potential. 
One can also see that the range of the nuclear potential is much larger 
in the square-well potential as compared to the Woods-Saxon potential. 
We have confirmed that these features remain the same 
even if we replace $e^{-iKr}$ in Eq. (\ref{bc1}) with 
$H_l^{(-)}(Kr)$ by taking into account the centrifugal and the Coulomb 
potentials in the inner region. 

\begin{figure} [tb]
\includegraphics[scale=0.6,clip]{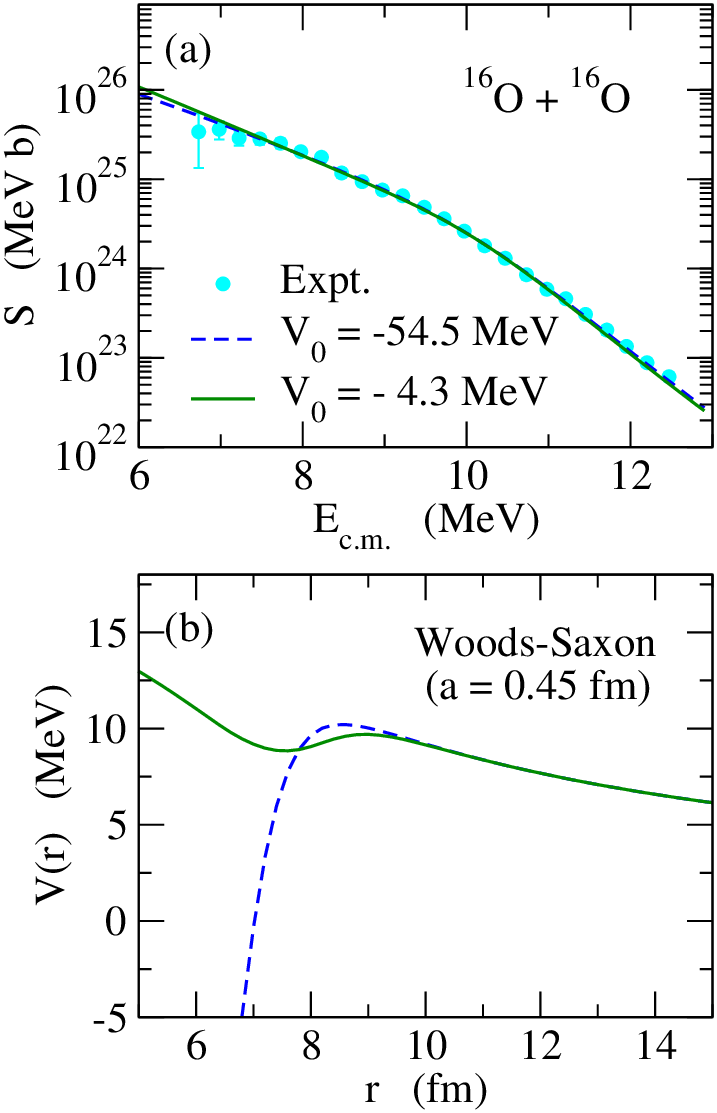}
\caption{
Same as Fig. 1, but with a shallow Woods-Saxon potential (the solid lines). 
The meaning of the dashed line is the same as in Fig. 1. 
}
\end{figure}

Because of continuous and discrete ambiguities of optical potentials  
\cite{BTW85,Hodgson1,Hodgson2,Cook80}, 
the parameters of the Woods-Saxon potential may not be determined 
uniquely. For instance, Fig. 2 shows the result with 
$V_0=-4.3$ MeV, $R=8.13$ fm, $a=0.45$ fm, $W_0=10$ MeV, $R_w=7.55$ fm, and 
$a_w=0.1$ fm. This potential is much shallower than the Woods-Saxon potential 
shown in Fig. 1, but still yields a comparable fit to the experimental 
data. That is, one can reproduce the data equally well by using either 
a deep potential with a small value of $R$ or a shallow potential with a 
larger value of $R$. Notice that the latter potential has a similar feature 
to the square-well potential shown in Fig. 1. 

For a Woods-Saxon potential, a change in the radius parameter 
can be compensated with a change in the depth parameter so that the 
height of the Coulomb barrier remains the same. 
In contrast, for the square-well potential, the potential changes abruptly 
at $r=R$, and the height of the Coulomb barrier is determined only by 
$R$. That is, the height is independent of $V_0$. 
The value of $R$ then cannot be too small, otherwise the Coulomb barrier 
is too high, considerably suppressing the 
astrophysical $S$-factor at $E\gtrsim 10$ MeV. 

\begin{figure} [tb]
\includegraphics[scale=0.6,clip]{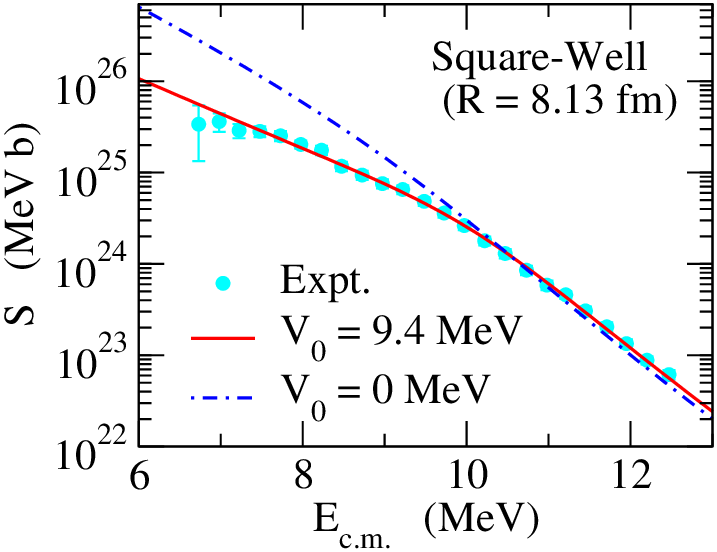}
\caption{
Comparison of the astrophysical $S$-factor for the 
$^{16}$O+$^{16}$O system obtained with two different depth parameters of the 
square-well potential. 
The solid line is the same as that in Fig. 1 and is obtained 
with $V_0=9.4$ MeV, while the dot-dashed line is obtained with 
$V_0=0$ MeV. The range parameter of the square-well potential is set 
to be $R=8.13$ fm for both cases. 
The experimental data are 
taken from Ref. \cite{Thomas86}. 
}
\end{figure}

In the square-well model of Michaud and Fowler, 
the range parameters for the real and the 
imaginary parts are set to be the same to each other 
\cite{BTW85,MF70,Fowler75}. 
A large value of $R$ for the 
real part then implies that the flux is absorbed from relatively large 
distances. 
In order to see this, 
Fig. 3 shows the result of the square-well potential with $V_0=0$ MeV, 
for which the inner region is classically allowed for the entire range 
of energy shown in the figure. 
One can observe that this calculation overestimates the 
astrophysical $S$-factor 
at energies below 10 MeV. 
Notice that, when the potential is shallow, the wave function in the 
inner region is largely damped if the incident energy is below $V_0$. 
Evidently, one has to make the potential shallow in order to 
reproduce the experimental data, which results in a hindrance of the absorption effect. 

\begin{figure} [tb]
\includegraphics[scale=0.6,clip]{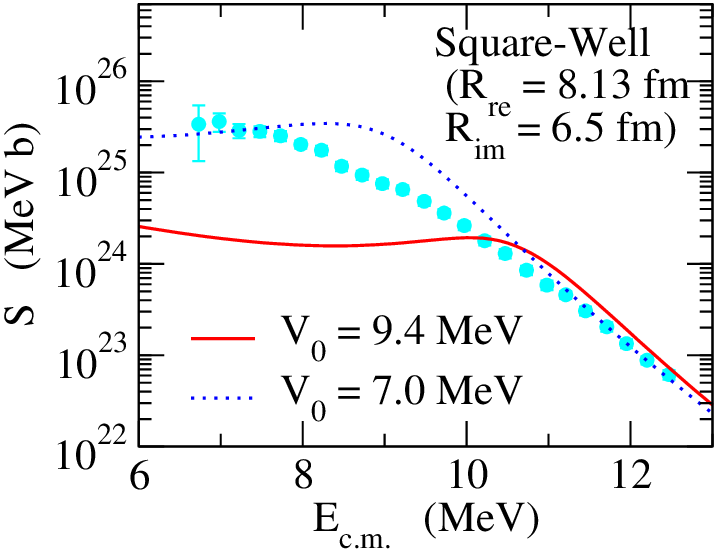}
\caption{
Same as Fig. 3, but for the case where the range parameter of the 
imaginary part of the square-well potential is set independently 
to that for the real part. 
Those are taken to be 8.13 fm and 6.5 fm for the real and the imaginary 
parts, respectively. 
The solid line is obtained 
with $V_0=9.4$ MeV, while the dotted line is obtained with 
$V_0=7.0$ MeV. 
The experimental data are 
taken from Ref. \cite{Thomas86}. 
}
\end{figure}

One would then expect that the depth of the square-well potential 
becomes deeper if the absorption 
range is shorter. This is indeed the case as is shown in Fig. 4, which 
is obtained by setting the range parameters for the real and the imaginary 
parts to be 8.13 and 6.5 fm, respectively. 
Notice that, in this case, 
the absorption does not start even if the relative motion penetrates 
through the barrier and reaches at $r=R$. 
To draw Fig. 4, 
we employ the boundary conditions of 
\begin{eqnarray}
u(r)&=&Te^{-iKr} ~~~~~~~~~~~~~~~~~~~~~~~~(r < R_i), \\
&=&Ae^{-i\tilde{K}r}+Be^{i\tilde{K}r}~~~~~~~(R_i\leq r < R), 
\end{eqnarray}
with $\tilde{K}=\sqrt{2\mu(E-V_0)/\hbar^2}$. 
The boundary condition for the outer region, $r\geq R$, remains the same 
as in Eq. (\ref{bc2}). 
The solid line in the figure is obtained with the same value of $V_0$ 
as in Fig. 3. Since the relative motion has further to penetrate 
the barrier before the absorption is effective, 
the astrophysical $S$-factor is largely underestimated. 
This is cured to some extent by deepening the potential depth, as is shown 
by the dotted line, which is obtained with $V_0=7.0$ MeV. 
The reproduction of the experimental data, however, is less satisfactory 
as compared to the solid line in Fig. 3. 
If the depth of the potential 
is further deepened, the astrophysical $S$-factors 
are overestimated as in the dot-dashed line in Fig. 3. 
Therefore, the choice of $R_i\neq R$ for the square-well 
model does not seem to be preferred, at least for the $^{16}$O+$^{16}$O system. 

\section{Summary}

We have investigated the origin of a shallow depth of a square-well 
potential for nuclear astrophysical reactions. 
We have argued that this is caused by the following two effects. Firstly, 
the square-well potential changes abruptly at the 
boundary, leading to a large radius parameter. 
Because of this, the absorption of the flux starts from relatively large distances. 
The potential depth then becomes shallow in order to hinder the absorption 
effect. It is important to notice that these are artifacts of a square-well 
potential, and a shallow depth has nothing to do with microscopic origins  
of a repulsive core in internuclear potentials, 
such as the Pauli principle effect \cite{Tamagaki65,Baye82}. 
Indeed, if one uses a Woods-Saxon potential, one can employ a more reasonable 
value for the radius and the depth parameters. 
This would imply that a care must be taken in interpreting the results 
of a square-well model and in extrapolating the results down 
to astrophysically relevant energies, even though the model is simple and 
convenient, and often provides a good fit to experimental data.

\section*{Acknowledgments}

We thank X.D. Tang for useful discussions. 
C.A.B. thanks the Graduate Program on
Physics for the Universe (GPPU) of Tohoku University for
financial support for his trip to Tohoku University, 
the U.S. NSF grant No 1415656, and the U.S. DOE grant No. DE-FG02-08ER41533.

\end{document}